\documentclass[%
 reprint,
superscriptaddress,
longbibliography,
 amsmath,amssymb,
 aps,
]{revtex4-1}

\usepackage[T1]{fontenc}
\usepackage{graphicx}
\usepackage{dcolumn}
\usepackage{bm}
\usepackage{amsmath}
\usepackage{lmodern}
\usepackage{mathtools}

\usepackage{xcolor}
\usepackage{graphicx}
\usepackage{dcolumn}
\usepackage{bm}
\usepackage[hidelinks]{hyperref}

\usepackage[normalem]{ulem}

\hypersetup{
    colorlinks,
    linkcolor= [rgb]{0.1804,0.1882,0.5725},
    citecolor=[rgb]{0.1804,0.1882,0.5725},
    urlcolor=[rgb]{0.1804,0.1882,0.5725}
}

\begin{document}

\title{Giant spin Meissner effect in a non-equilibrium exciton-polariton gas}

\author{M.\,Kr\'ol}
\email{Mateusz.Krol@fuw.edu.pl}
\author{R.\,Mirek}
\author{D.\,Stephan}
\author{K.\,Lekenta}
\author{J.-G.\,Rousset}
\author{W.\,Pacuski}
\affiliation{Institute of Experimental Physics, Faculty of Physics, University of Warsaw, ul.~Pasteura 5, PL-02-093 Warsaw, Poland}
\author{A.\,V.\,Kavokin}
\affiliation{Institute of Natural Sciences, Westlake University, No.18, Shilongshan Road, Cloud Town, Xihu District, Hangzhou, China }
\affiliation{National University of Science and Technology "MISIS", 4, Leninskiy pr., 119049, Moscow, Russia}
\affiliation{Spin Optics Laboratory, St. Petersburg State University, Ul'anovskaya 1, Peterhof, St. Petersburg 198504, Russia}
\author{M.\,Matuszewski}
\affiliation{Institute of Physics, Polish Academy of Sciences, al.~Lotnik\'{o}w 32/46, PL-02-668 Warsaw, Poland}
\author{J.\,Szczytko}
\author{B.\,Pi\k{e}tka}
\email{Barbara.Pietka@fuw.edu.pl}
\affiliation{Institute of Experimental Physics, Faculty of Physics, University of Warsaw, ul.~Pasteura 5, PL-02-093 Warsaw, Poland}

\begin{abstract}
The suppression of Zeeman energy splitting due to spin-dependent interactions (the spin Meissner effect) was predicted to occur within a Bose-Einstein condensate.
 We report a clear observation of this effect in semimagnetic microcavities which exhibit a giant Zeeman energy splitting between two spin-polarised polariton states as high as 2\,meV, and demonstrate that a partial suppression of energy difference occurs already in the uncondensed phase in a striking similarity to the up-critical superconductors in the fluctuation dominated regime. These observations are explained quantitatively by a kinetic model accounting for both the condensed and uncondensed polaritons and taking into account the non-equilibrium character of the system.

\end{abstract}

\maketitle


One of the defining properties of a superconductor is the expulsion of external magnetic field from its interior by surface currents appearing on the boundary. This phenomenon, known as the Meissner effect~\cite{Annet}, has its analogue in a neutral bosonic condensate with spin. In this case, the minimization of free energy in the presence of magnetic field leads to a complete screening of the Zeeman splitting by the interactions between two spin subsystems\,\cite{Rubo_2006}. Indeed, the imbalance between populations of spin components results in an effective magnetic field that exactly cancels the external magnetic field in the ground state of the system. This exact compensation of the external magnetic field effect is expected to occur at the fields up to some critical value dependent on the $g$-factor of bosonic quasiparticles, the interaction constants of spin-parallel and spin-antiparallel bosons and the concentration of bosons.

Exciton-polaritons in semiconductor microcavities constitute an example of spinor bosonic quasiparticles: a mixture of matter excitation in semiconductor quantum wells (excitons) and photons confined in a cavity structure~\cite{Kasprzak,Kavokin_Microcavities}. 
The excitons that can couple to light have a $\pm$1 spin degeneracy, and therefore couple respectively to photons of opposite chirality, $\sigma^{\pm}$, forming two subsystems of exciton-polaritons of opposite spins. They can be considered as a bosonic quasiparticles with a 1/2 pseudospin. Most importantly, the interactions between polaritons are spin dependent~\cite{Tassone_PRB1999,Martin_PRL2002,Takemura_NatPhys2014,Deveaud_2016,Sun_NatPhys2017}. Typically, polaritons with the same spin projection on the quantum well axis strongly repel, while polaritons with opposite spin projection weakly attract \cite{Fernandez-Rossier_PRB1996,Vladimirova_PRB2010}. This interaction constant difference together with population imbalance are responsible for the appearance of the spin Meissner effect in a polariton condensate. On one hand, due to the Zeeman effect and thermalisation processes, polaritons tend to orient their spins in external magnetic fields, on the other hand, polaritons try to minimize the free energy due to the strong repulsion of polaritons with the same spin. Therefore there is an interplay between spin polarization induced by external magnetic field and polariton-polariton interaction, what leads to suppression of spin splitting, that acts effectively as the expulsion of magnetic field from the superconductor interior.

The theoretical prediction of the spin Meissner effect in polariton condensate presented in Ref.\,\citep{Rubo_2006} was followed by multiple experimental works \cite{Larionov_PRL2010,Kulakovskii_PRB2012,Fischer_PRL2014,Sturm_PRB2015, Walker_PRL2011}. Suppression of the Zeeman splitting \cite{Larionov_PRL2010}, even up to sign reversal \cite{Fischer_PRL2014} were reported, but were accompanied by an unexpected linear polarization behavior and deviations from the predictions of the thermal equilibrium model in the polarization splitting dependence on the field \cite{Chernenko_2016,Chernenko_Semic2018}. Most of the authors explain deviations from the theory due to the non-equilibrium polariton dynamics \cite{Sturm_PRB2015,Kulakovskii_PRB2012,Fischer_PRL2014}, whilst the stationary regime close to the thermal equilibrium in polariton condensates is achieved only in the last generation of samples \cite{Sun_PRL2017,Caputo_NatMater2018}. In electrically-driven polariton lasers, suppression of the Zeeman splitting was  observed in the polariton-lasing regime \cite{Schneider_2013}, but in other realization of a similar experiment a circular polarization of the emission was attributed to the Zeeman splitting\,\cite{Bhattacharya_PRL2013}.

\begin{figure}
\centering
\includegraphics[width=.45\textwidth]{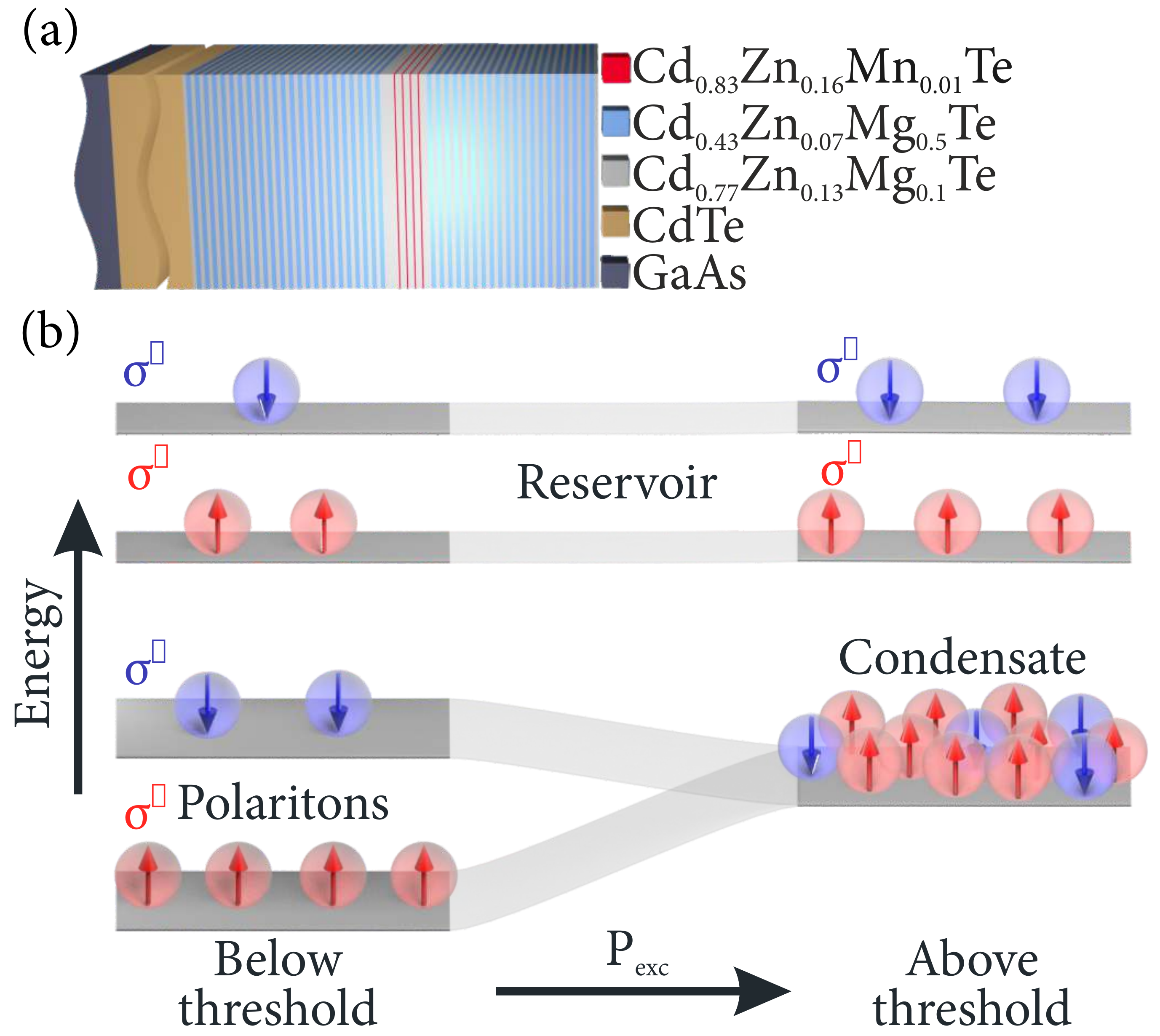}
\caption{(a) Scheme of the investigated microcavity with 20 (22) top (bottom) CdZnMgTe DBR pairs and four quantum wells doped with magnetic Mn\textsuperscript{2+} ions. (b) Energy levels in external magnetic field with increasing excitation power. Above the threshold the condensate Zeeman splitting is suppressed, while the reservoir splitting is still present.}
\label{im:Fig0}
\end{figure}

The most important parameter that up to now made difficult the proper experimental observation of those effects was the small energy splitting of exciton-polaritons of opposite spins (weak Zeeman effect). Using semimagnetic semiconductors allow us to overcome this limitation. Semimagnetic materials were also  considered in theoretical works investigating the spin Meissner effect \cite{Shelykh_PRB2009}. The theory assuming the minimization of free energy of the condensate energy for a fixed position of magnetic ions in the system predicted a linear polarisation of a polariton condensate in equilibrium and an increasing degree of circular polarization with increasing magnetic field. Our first results in CdMnTe-based semimagnetic semiconductor microcavities showed an increase of the degree of circular polarization of the condensate with magnetic field \cite{Rousset_PRB2017, Krol_SciRep2018} but did not show a strong evidence of quenching of the Zeeman splitting. Here, we demonstrate that in specially designed system supporting a semimagnetic polariton condensate with giant spin-splitting \cite{Mirek_PRB2017} and careful choice of excitation parameters a suppression of the Zeeman splitting around the condensation threshold is observed, which we believe is a clear signature of the spin Meissner effect predicted in Ref.\,\citep{Rubo_2006}. The magnitude of the quenching is as high as 2\,meV at 6\,T. We extend also the theoretical model provided there to account for a non-equilibrium character of the polariton condensate. 

Moreover, we present an experimental evidence of a new regime of a polariton gas, which we call the reservoir-dominated regime, in analogy to the fluctuation-dominated regime of fluctuating superconductors. The fluctuation-dominated regime is characterized by the occurrence of partial Meissner effect~\cite{Varlamov_RevModPhys2018}. We observe an analogous phenomenon in a polariton gas, with a partial spin Meissner effect below the laser threshold. This is a very new observation as compared to previous works where the features of spin Meissner effect were detected only above the polariton lasing threshold. The observation of this regime is of a fundamental importance as it shows the similarity of our bosonic system and fluctuating superconductors \cite{Varlamov_RevModPhys2018}. The reservoir-dominated phase appears in our system as it offers a supplementary mechanism of fluctuations: the magnetization fluctuations in the semimagnetic structure that affect the potential seen by exciton-polaritons. This specific feature of a semimagnetic cavity makes it an excellent test-bed for revealing fluctuation effects in a bosonic polariton gas.

\begin{figure}
\centering
\includegraphics[width=.45\textwidth]{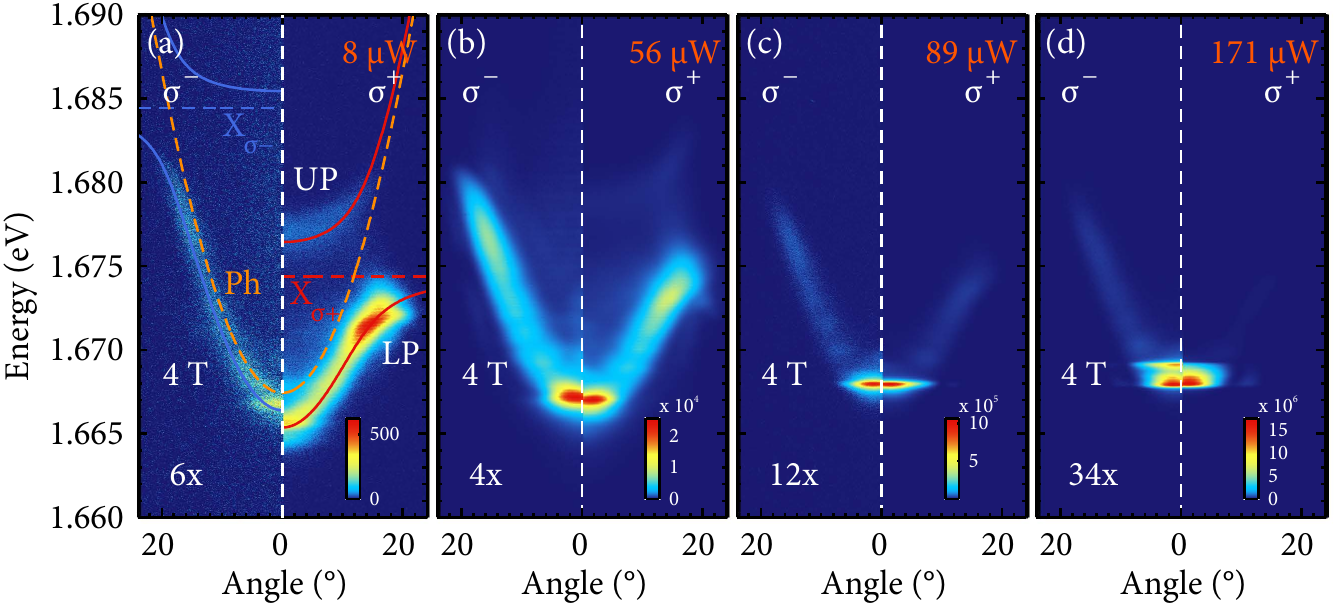}
\caption{Experimental data illustrating the semimagnetic condensate formation upon increasing excitation power at 4~T. The system is excited with a linearly polarized ps-pulsed laser at the energy of the first Bragg minimum of the structure at the high energy side (approx. 1.746\,eV). The photoluminescence spectra are time integrated and are angularly resolved with the angle corresponding to the polariton in-plane momentum. The dashed lines represent the energy modes of the uncoupled exciton and photon system, while polariton modes are marked by red and blue solid lines, corresponding to $\sigma^+$ and $\sigma^-$ polarization detection, respectively.}
\label{im:Fig1}
\end{figure}

We investigated a non-magnetic CdTe-based microcavity with quantum well doped with magnetic Mn ions presented schematically in Fig.\,\ref{im:Fig0}(a) and described in more detail in \cite{Rousset_JCrystGrowth2013,Rousset_APL2015}.
In the CdMgZnTe microcavity structure and \mbox{CdMnTe} semimagnetic material of the quantum well, the \textit{s},\textit{p}-\textit{d} exchange interaction between localized \textit{d}-shell electrons of the magnetic ions and the \textit{s}-shell electrons and \textit{p}-shell holes leads to magneto-optical effects such as giant Faraday rotation or giant Zeeman splitting \cite{Gaj_1978}. The scheme of the  structure is illustrated in Fig.\,\ref{im:Fig0}(a). In Ref.~\cite{Mirek_PRB2017} we demonstrated a giant Zeeman splitting of exciton-polaritons in the same microcavity structure, where magnetic ions are present only in  quantum wells, affecting only the excitonic component of the polariton state. We have shown that the external magnetic field can induce condensation by reducing the condensation threshold power due to the decrease of the available density of states up to a factor of two which reduces the condensation threshold \cite{Rousset_PRB2017}. Moreover, we demonstrated the creation of a spin multicomponent condensate and the possibility to tune it smoothly to a single component condensate by an external parameter, i.e. excitation power and/or magnetic field \cite{Krol_SciRep2018}.

\begin{figure*}
\centering
\includegraphics[width=.95\textwidth]{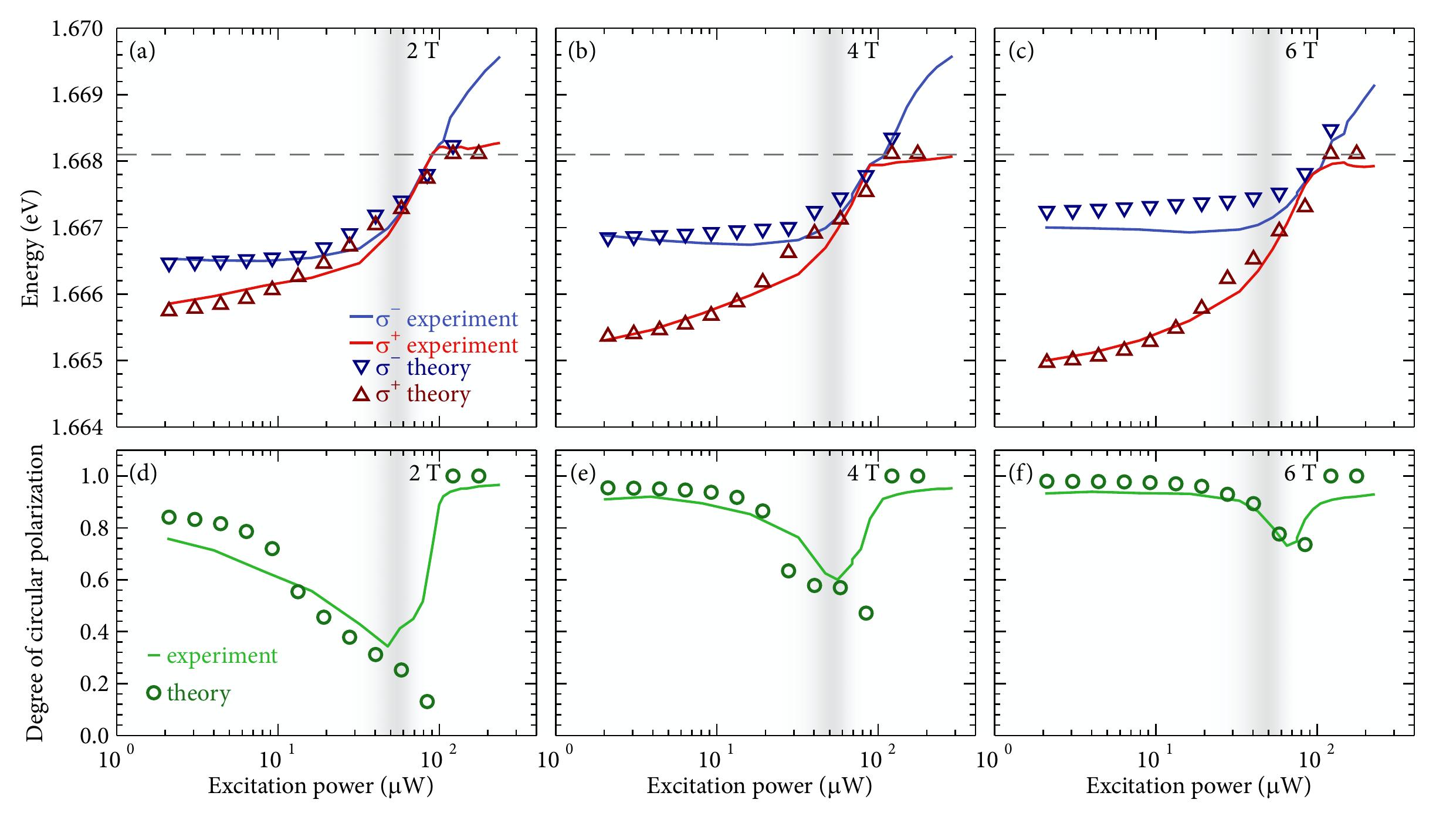}
\caption{The energy (a--c) and degree of circular polarization (DOCP) (d--f) of the signal at the bottom of lower polariton branch versus excitation power in magnetic fields: 2\,T, 4\,T and 6\,T. The energy difference between $\sigma^+$ and $\sigma^-$ polariton states decreases with the pumping power with complete suppression of the splitting close to the condensation threshold, where the minimum of the DOCP occurs. Experimental results are marked with solid lines with fitted model represented with points. The condensation threshold, marked by the shadow, is 60\,$\mu$W at 0 T and is slightly decreasing in magnetic field.}
\label{im:Fig2}
\end{figure*}

The results presented here demonstrate a new approach to the spin-Meissner effect of semimagnetic exciton-polariton spinor condensates, where the magnetic interactions play a crucial role. We create a polariton condensate at several values of the external magnetic field by increasing the excitation power, which is schematically shown in Fig.\,\ref{im:Fig0}(b). At fixed magnetic field and below the condensation threshold [Fig.\,\ref{im:Fig0}(b)] the system exhibits giant Zeeman energy splittings both in the polariton state (along the lower polariton branch) and in the excitonic reservoir (that can be traced at high emission angles, where polaritons are mostly excitonic). With increasing pumping power a quenching of the Zeeman splitting at the polariton mode followed by the formation of polariton condensate is observed. The uncondensed polaritons that form the reservoir still exhibit a giant Zeeman splitting. Measurements were performed for negative exciton--photon detuning of $-12$\,meV with Rabi splitting equal to 7.7\,meV without the external magnetic field.
 Fig.\,\ref{im:Fig1} shows the evolution  of the experimental emission spectra of semimagnetic polaritons at the magnetic field of 4\,T with increasing excitation power. The spectra for other values of magnetic field ranging from 0\,T to 6\,T are provided in SI \cite{SI}.
As the excitation power increases we observe the polariton condensation at the bottom of the lower polariton branch that is accompanied by a quenching of the energy difference between the counter-polarized signal (in $\sigma^+$ and $\sigma^-$ polarization detection). Uncondensed polaritons, described within a two-mode coupling model \cite{Weisbuch}, marked by solid and dashed lines in Fig.\,\ref{im:Fig1}, show Zeeman energy splitting as anticipated in the linear regime \cite{Pietka_PRB2015, Mirek_PRB2017}. The polariton condensation at the bottom of the lower polariton branch is also revealed by a nonlinear increase of the emission intensity, linewidth narrowing and energy blueshift due to the interactions present in the system, which is illustrated in detail in the SI \cite{SI}. The condensation threshold can be found for each magnetic field, and it equals 60\,$\mu$W at 0\,T. It slightly decreases in magnetic field, in agreement with the previous studies \cite{Rousset_PRB2017}.

In Fig.~\ref{im:Fig2}(a--c) we plot the energy of the emission lines as a function of the excitation power for both predominantly $\sigma^+$ and $\sigma^-$ polarized states. We observe that with an increase of the excitation power the energy difference between $\sigma^+$ and $\sigma^-$ states is decreasing, moreover, at the threshold the energy difference becomes smaller than the linewidth and cannot be resolved. With a further increase of the excitation power we observe that  both spin components of the condensate have the same energy, which increases due to the interactions in the system. Fig.~\ref{im:Fig3}(a) illustrates the energy difference between the two circularly polarized components of the condensate and Fig.\,\ref{im:Fig3}(b) shows the accompanying splitting of uncondensed polaritons that represents the Zeeman splitting of excitonic reservoir. The threshold power for each magnetic field is marked by the crosses. We believe that reduction of the Zeeman splitting of the condensate down to zero is a clear signature of the spin Meissner effect. It shows the importance of spin dependent interactions within the condensate. In addition, a partial reduction of Zeeman splitting with increasing pumping power is observed for polaritons gas well below the condensation threshold, a behaviour which is not predicted by existing models.

To explain the above results
--- indicating that reduction of giant Zeeman splitting can be clearly recognized even below the
condensation threshold, which is visible in Fig.\,\ref{im:Fig3}(a) --- we introduce a phenomenological model of the polariton reservoir and condensate dynamics. Polariton kinetics are accounted for in a similar way as in several previous works~{\cite{Kavokin_PRL2004,Iorsh_PRB2012,Bhattacharya_PRB2016}, %
where spinor polariton condensates were investigated. Additionally, we describe the spin Meissner effect by adapting the description formulated in Ref.~\citep{Rubo_2006,Shelykh_SuperlatticesMicrostruct2007} but generalised to the system in a non-equilibrium state. To account for both the reservoir dynamics and condensate formation, as well as the photon lasing at high pumping powers, we formulate the equations for the reservoir density $n_{\rm R}(t)$ and its average spin ${\bf S}_{\rm R}(t)$, condensate density $n_{\rm c}(t)$, assuming that the spin of the condensate always corresponds to minimum of energy ${\bf S}_{\rm c}(t)={\bf S}^{\rm min}_{\rm c}(t)$, and the lasing mode occupation $n_{\rm L}(t)$
\begin{align*}
\frac{\partial n_{\rm R}}{\partial t} &= P-\gamma_{\rm R} n_{\rm R}-R_{\rm sc} n_{\rm R} n_{\rm c}-R_{\rm L} n_{\rm R} n_{\rm L}, \\
\frac{\partial {\bf S}_{\rm R}}{\partial t} &= -\gamma_{\rm rel} \left({\bf S}_{\rm R}-{\bf S}_{\rm R}^{\rm min}\right)-R_{\rm sc} n_{\rm c} {\bf S}_{\rm R}-\gamma_{\rm R} {\bf S}_{\rm R}, \\
\frac{\partial n_{\rm c}}{\partial t} &= R_{\rm sc} n_{\rm R} n_{\rm c}-\gamma_{\rm c} n_{\rm c}, \\
\frac{\partial n_{\rm L}}{\partial t} &= R_{\rm L} n_{\rm R} n_{\rm L}-\gamma_{\rm L} n_{\rm L},
\end{align*}
where $P$ is the pumping rate, $\gamma_{\rm R, c}$ are excitonic and photonic decay rates, $R_{\rm sc} \approx R_{0} + R_1 E_{c}$ is the reservoir-condensate scattering rate, approximately linearly dependent on the condensate energy, $R_{\rm L}$ is the rate of scattering into the lasing mode, $\gamma_{\rm rel} = \gamma_{\rm phonon} + \Gamma_{\rm int} n_{\rm R}$ is the reservoir relaxation rate, which accounts for the phonon mediated relaxation $\gamma_{\rm phonon}$ and interaction mediated relaxation $\Gamma_{\rm int}$. 
The reservoir and condensate spins that minimize the energy of the system  ${\bf S}_{\rm R,c}^{\rm min}$ are calculated as in \citep{Rubo_2006,Shelykh_SuperlatticesMicrostruct2007}, from the minimization of the free energy
\begin{equation}\label{En}
F=-2\Omega_z S^z+2\Omega_x S^x+\frac{\alpha}{2} \left(n_+^2 + n_-^2\right) - \mu n,
\end{equation}
where $\Omega_z$ is the excitonic giant Zeeman splitting appearing due to the magnetic field parallel to the sample growth axis,  $\Omega_x$ is the linear polarization splitting due to the sample anisotropy or from the residual transverse magnetic field, $S^{x,z}=S_R^{x,z} + X S_c^{x,z}$ is the $z$ component of the total excitonic spin, where $X$ is the Hopfield coefficient of condensed polaritons, $n^\pm=n_{\rm R}^\pm + X n_{\rm c}^\pm$
where $n_{\rm R, c}^{\pm}=(n_{\rm R, c}/2)  \pm S^z_{\rm R,c}$, $\alpha$ is the parallel spin interaction coefficient, and we neglect the interaction between polaritons with antiparallel $z$-spin projection. Theoretical and experimental estimates indicate that such interaction is much weaker than in the parallel spin configuration \cite{Ciuti_PRB1998}, and we found that the results presented in our work can be well explained neglecting polariton-polariton interactions in the antiparallel spin configuration.

\begin{figure}
\centering
\includegraphics[width=.4\textwidth]{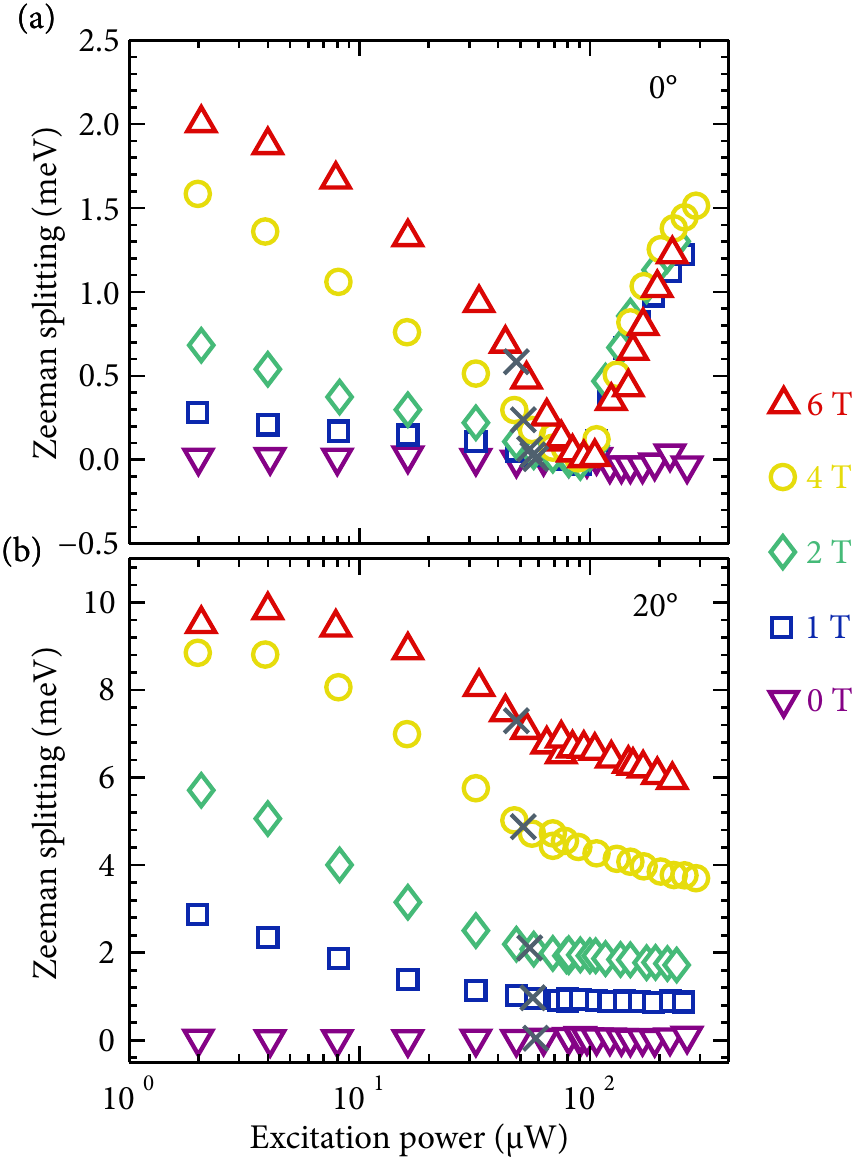}
\caption{Energy splitting between two spin components of (a) exciton-polariton condensate at bottom of lower polariton branch (b) non-condensed polaritons reservoir detected at 20$^\circ$ angle. The condensation threshold power at each magnetic field is marked with gray cross.
}
\label{im:Fig3}
\end{figure}

According to the previous theories, the spin Meissner effect is expected to be accompanied by a specific change of spin polarization. In the presence of the external magnetic field, the minimum of the energy is achieved when Zeeman splitting is compensated by the polariton--polariton interactions. Effective screening of the external magnetic field originates from the accumulation of polaritons with spin parallel to the magnetic field, which creates an effective counteracting magnetic field. In the case of perfect equilibrium at zero temperature, the external field is completely compensated for in the ground polariton state\,\cite{Rubo_2006}. If the full equilibrium is not achieved, we can expect a partial reduction of the energy splitting. This effect is observed below the condensation threshold (vertical shaded line in Fig.~\ref{im:Fig2}), when the two emission lines are still well separated, but exhibit a partial reduction of the splitting. The reduction of Zeeman splitting at the pump power below the polariton laser threshold is a peculiar demonstration of the collective behavior of exciton-polaritons prior to formation of extended condensates. A similar, partial reduction of the external magnetic field is observed in a number of up-critical superconductors in the fluctuation dominated regime~\cite{Larkin}.

At very high excitation powers the two counter polarized signals split again, which we attribute to the loss of strong coupling and the transition to photon lasing in the dominant polarization ($\sigma^+$). The energy of the $\sigma^+$ component at this very high excitation power is fixed at the photon energy, however the energy of the $\sigma^-$ component is still increasing even though this state is not much populated. This suggests that the strong coupling is retained for excitons polarized opposite to magnetic ions  \cite{Ballarini_APL2007}. The accompanying change of the emission intensity is illustrated in the SI \cite{SI}. 

The variation of the degree of circular polarization (DOCP) with the excitation power at different magnetic fields is presented in Fig.~\ref{im:Fig2}(d--f). We define DOCP as: $\rho=\frac{I_{\sigma^+}-I_{\sigma^-}}{I_{\sigma^+}+I_{\sigma^-}}$, where $I_{\sigma^{\pm}}$ denote the emission intensities 
of the most populated state detected in corresponding circular polarizations. Due to the large Zeeman splitting, the emission is almost fully circularly polarized in most cases, as shown in Fig.~\ref{im:Fig2}(d--f). However, in the vicinity of the condensation threshold and slightly below it, in the regime where the compensation of the external magnetic field occurs, the degree of circular polarization decreases. This reduction of the degree of circular polarization is due to the presence of a nonzero linear polarization splitting $\Omega_x$, and the spin Meissner effect which reduces the effective magnetic field in $z$ direction, cf.\,Eq.\,(\ref{En}). Above the photon lasing threshold, where the second splitting occurs, the emission builds up the spin polarization very rapidly, and the system becomes fully polarized. 

We would like to comment also on the additional effect that is important for exciton-polariton condensates in semimagnetic structures, ie. the reduction of the exciton Zeeman splitting due to the depolarization of the Mn ion subsystem by a high number of free particles created by high power and non-resonant optical excitation \cite{Koenig,Golnik_2004} (already discussed in \cite{Krol_SciRep2018}). This effect is visible at high emission angles, where we can trace out the almost pure excitonic component of the polariton state. The cross-section of the polariton emission at high angles (20\,deg.) is illustrated in the SI \cite{SI}. Fig.~\ref{im:Fig3}(b) illustrates the Zeeman splitting of exciton-like exciton--polaritons that is observed in our structures at given excitation powers. It is reduced due to the Mn-depolarization effects but it is clearly visible also at the polariton lasing threshold  (marked with grey cross). At the highest used excitation powers the Zeeman splitting is as high as 6\,meV at 6\,T and is always non-zero for lower field values. Therefore, the effect of  the almost complete quenching of the Zeeman splitting of the polariton condensate cannot be attributed to heating.

\begin{figure}
\centering
\includegraphics[width=.49\textwidth]{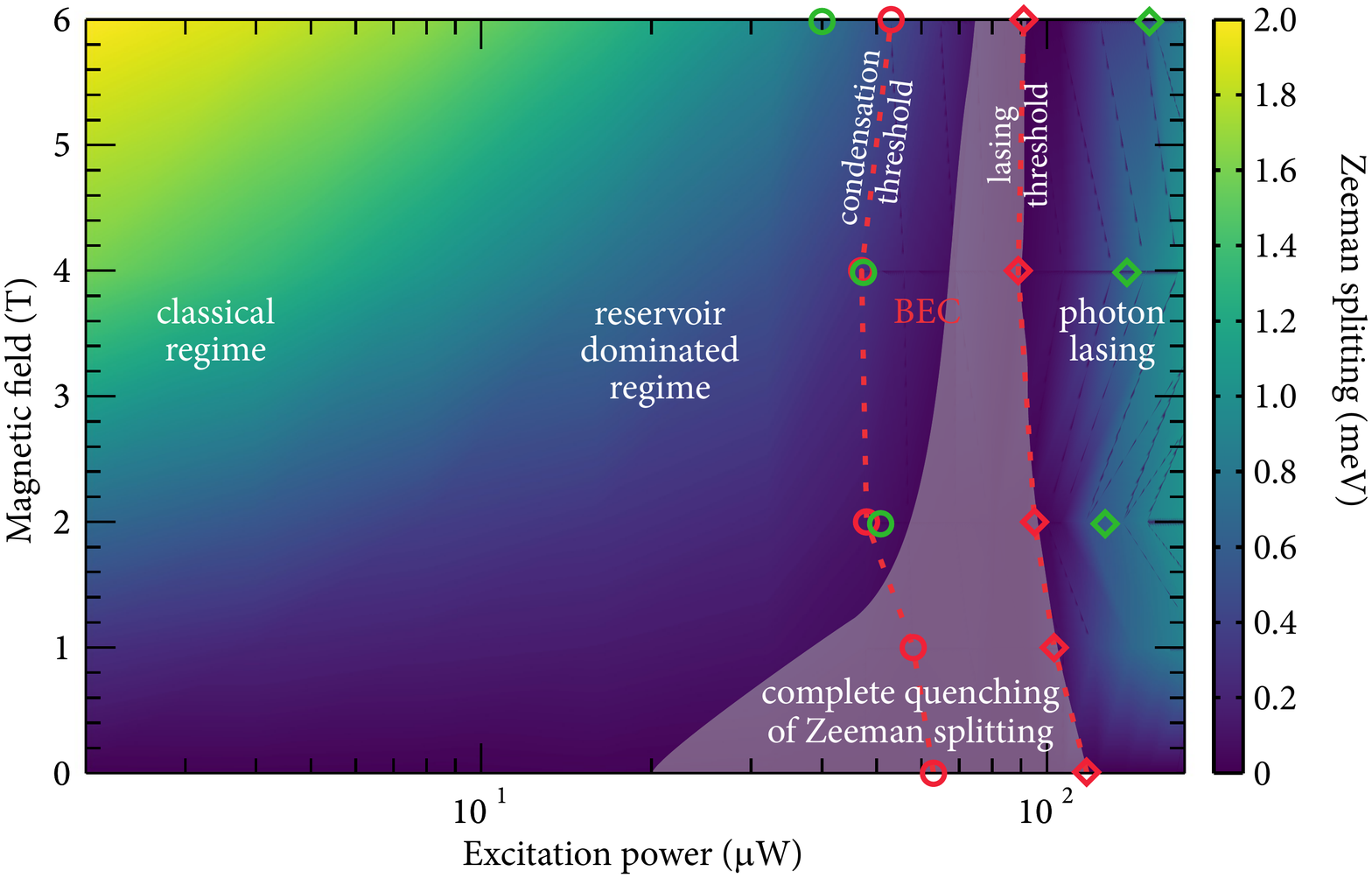}
\caption{Phase diagram of an exciton-polariton gas in a semimagnetic microcavity in external magnetic field. Complete suppression of Zeeman splitting is observed already below condensation threshold, in the reservoir-dominated regime. Threshold powers for condensation (photon lasing) are marked with circles (diamonds). The experimental and theoretical values are shown in red and green color, respectively.}
\label{im:Fig4}
\end{figure}

Obtained results are summarized by the phase diagram shown in Fig.\,\ref{im:Fig4}. In the classical regime, at low polariton densities, the Zeeman splitting increases with the magnetic field. Complete quenching of the Zeeman splitting occurs already below the condensation threshold and further on, in the condensed phase it is supported to higher magnetic fields due to the spin Meissner effect. For higher pump intensities the photonic lasing occurs. This behaviour is fully confirmed by our theoretical model. Note that the apparent discrepancy between theoretical and experimental photon lasing thresholds in Fig.\,\ref{im:Fig4} is due to the fact that in the experiment there is no clear threshold and the transition is smooth, as clearly visible in  Figs.~\ref{im:Fig2}(d)-(f). We attribute this discrepancy either to the disorder in the sample (significant in CdTe-based microcavities \cite{Kasprzak,Lagoudakis_2008,Krizhanovskii_PRB2009}) resulting in different lasing thresholds in different parts of the sample, or to the treatment of the reservoir in the model by a single equation, which does not take into account the full complexity of reservoir.

In summary, the exciton-polariton energy splitting in magnetic field is governed by the competition of two counteracting effects: polariton-polariton interactions and Zeeman effect. In the equilibrium spin Meissner effect these two contributions are equal which is manifested as a reduction of the energy splitting between two states of opposite polarization. We demonstrate that in semimagnetic microcavities, which are characterized by a giant Zeeman splitting of excitons larger than the emission linewidth, this effect is visible as a gradual reduction of the exciton--polariton Zeeman splitting even in the uncondensed state, when the system is far from thermal equilibrium. Above the condensation threshold, we observe that the energy splitting in the condensate is almost completely quenched in accordance with the equilibrium theory predictions. 

Finally, the similarity of the Spin Meissner effect below the polariton lasing threshold and the partial Meissner effect in the fluctuation dominated regime in superconductors is phenomenological: while in superconductors the effect is due to virtual Cooper pairs, in a polariton system it is governed by the spin-dependent polariton interactions with the exciton reservoir.

This work was supported by the National Science Center, Poland under projects 2014/13/N/ST3/03763, 2015/16/T/ST3/00506, 2015/18/E/ST3/00558, 2015/18/E/ST3/00559, 2015/17/B/ST3/02273, 2017/27/B/ST3/00271 and 2017/01/X/ST3/01861. This study was carried out with the use of CePT, CeZaMat and NLTK infrastructures financed by the European Union - the European Regional Development Fund. Scientific work co-financed from the Ministry of Higher Education budget for education as a research project "Diamentowy Grant": 0010/DIA/2016/45 and 0109/DIA/2015/44. AK acknowledges the support from the joint Russian-Greek project 'Polisimulator' supported by the Ministry of Education and Science of The Russian Federation (Project No. RFMEFI61617X0085). AK acknowledges the support of the Ministry of Education and Science of the Russian Federation in the framework of Megagrant No. 14.Y26.31.0027.


%

\end{document}


\begin{abstract}

	\end{abstract}

\author{M.\,Kr\'ol}
\email{Mateusz.Krol@fuw.edu.pl}
\author{R.\,Mirek}
\author{D.\,Stephan}
\author{K.\,Lekenta}
\author{J.-G.\,Rousset}
\author{W.\,Pacuski}
\affiliation{Institute of Experimental Physics, Faculty of Physics,
University of Warsaw, ul.~Pasteura 5, PL-02-093 Warsaw, Poland}
\author{A.\,V.\,Kavokin}
\affiliation{Institute of Natural Sciences, Westlake University, No.18,
Shilongshan Road, Cloud Town, Xihu District, Hangzhou, China }
\affiliation{National University of Science and Technology "MISIS", 4,
Leninskiy pr., 119049, Moscow, Russia}
\affiliation{Spin Optics Laboratory, St. Petersburg State University,
Ul'anovskaya 1, Peterhof, St. Petersburg 198504, Russia}
\author{M.\,Matuszewski}
\affiliation{Institute of Physics, Polish Academy of Sciences,
al.~Lotnik\'{o}w 32/46, PL-02-668 Warsaw, Poland}
\author{J.\,Szczytko}
\author{B.\,Pi\k{e}tka}
\email{Barbara.Pietka@fuw.edu.pl}
\affiliation{Institute of Experimental Physics, Faculty of Physics,
University of Warsaw, ul.~Pasteura 5, PL-02-093 Warsaw, Poland}


\title{Supplementary information for Giant spin Meissner effect in non-equilibrium exciton-polariton
condensate}
	\maketitle
	
\section*{Condensate emission energy, intensity and linewidth}
\begin{figure*}[h]
\centering
\includegraphics[width=.75\paperwidth]{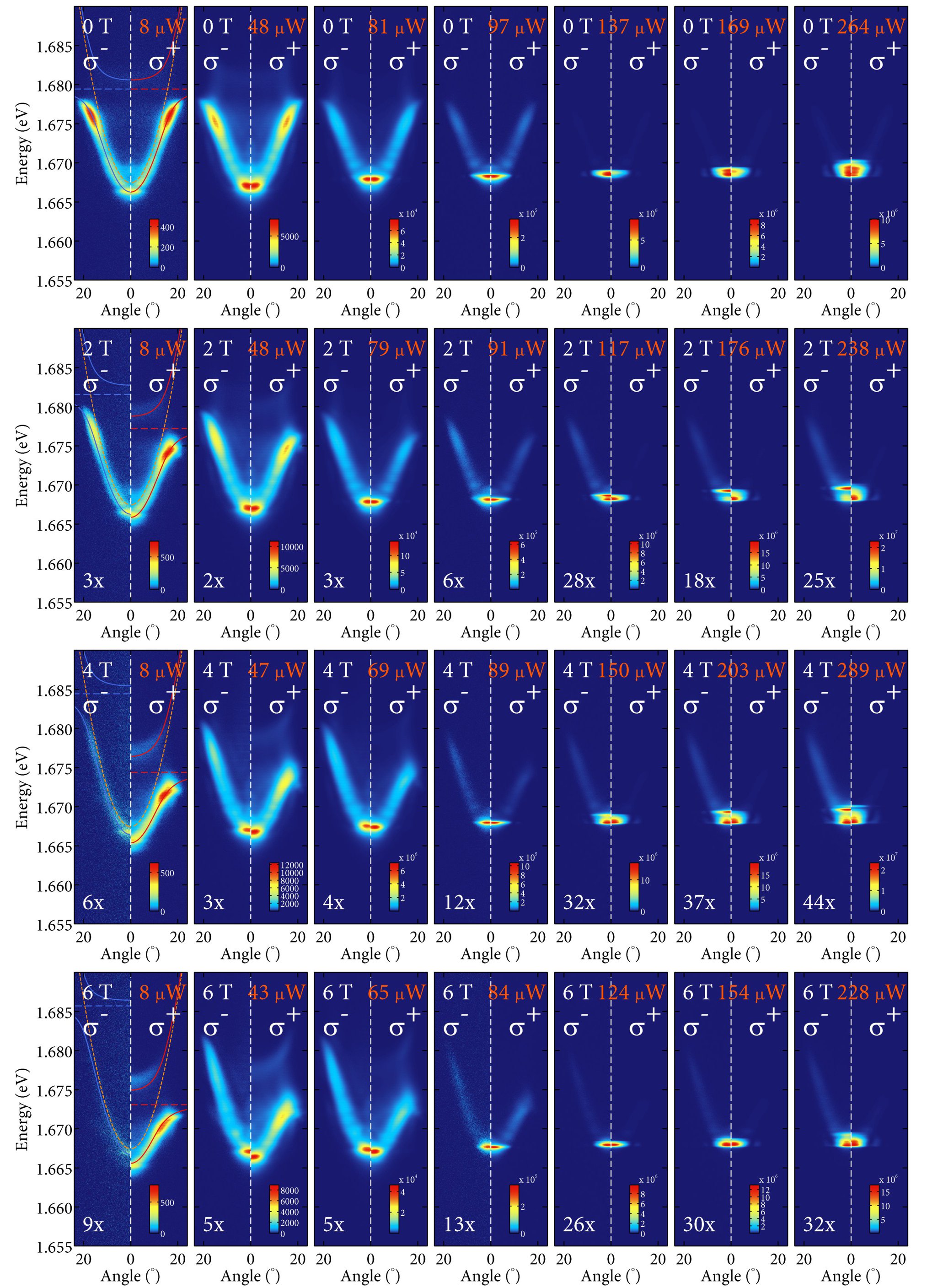}
\caption{Experimental data illustrating the condensate formation upon increasing excitation power at 0~T, 2~T, 4~T and 6~T.}
\label{im:Fig1}
\end{figure*}

Upon increasing excitation power we observe the polariton condensation at the bottom of lower polariton branch. The as measured angle-resolved emission spectra for different values of magnetic field are illustrated in Fig.~\ref{im:Fig1}. 

\begin{figure*}[h]
\centering
\includegraphics[width=.55\paperwidth]{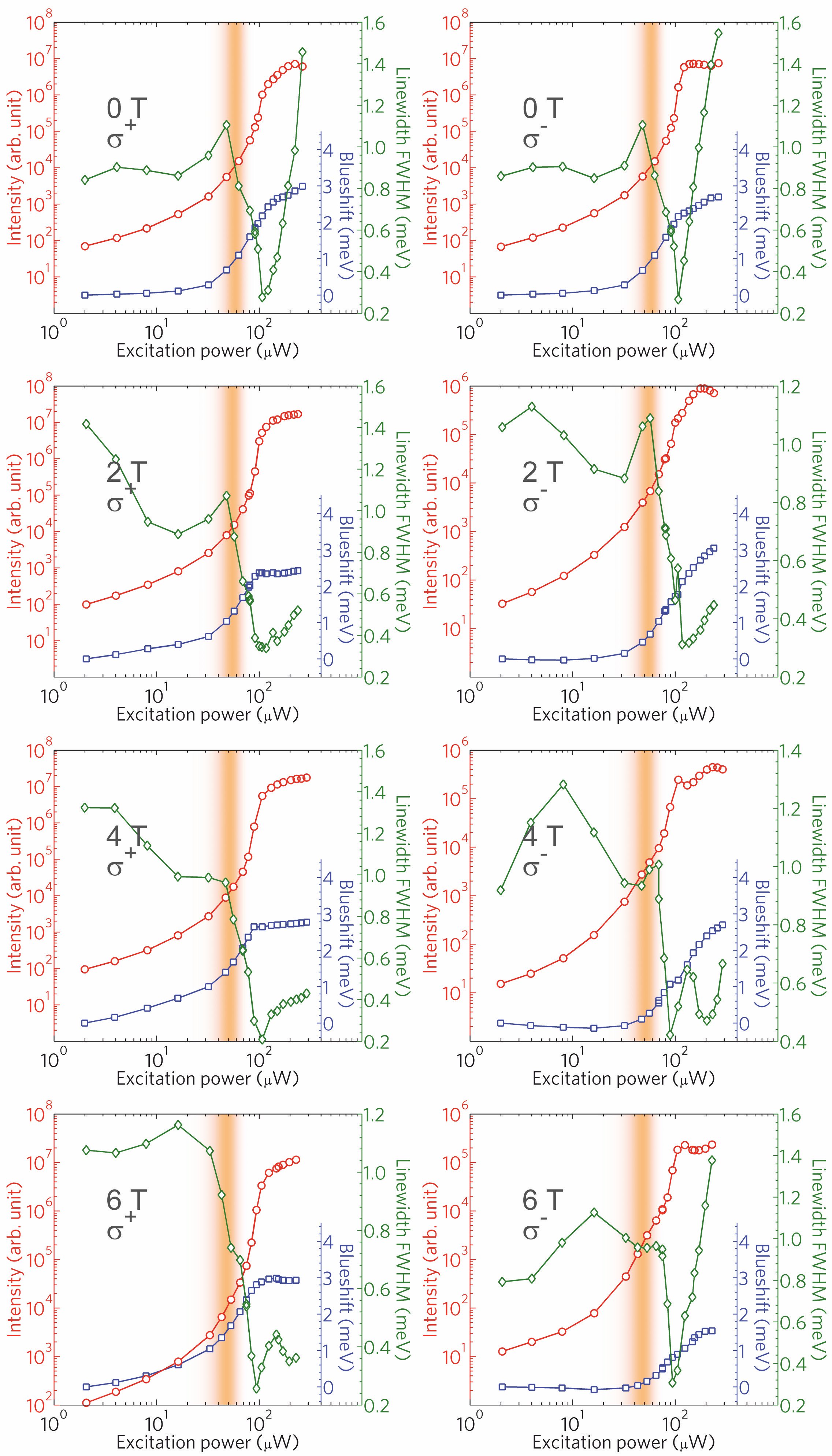}
\caption{The emission intensity, linewidth and energy shift of the bottom of lower polariton branch. The condensation threshold is determined to 60~$\upmu$W at 0~T.}
\label{im:Fig2}
\end{figure*}

The condensation at the bottom of lower polariton branch is revealed by the non-linear increase of the emission intensity, linewidth narrowing and energy blue shift due to the interactions present in the system, what is illustrated in Fig.~\ref{im:Fig2}.  We assign the condensation threshold to the power at which a significant linewidth narrowing occurs.

\begin{figure*}[h]
\centering
\includegraphics[width=.4\paperwidth]{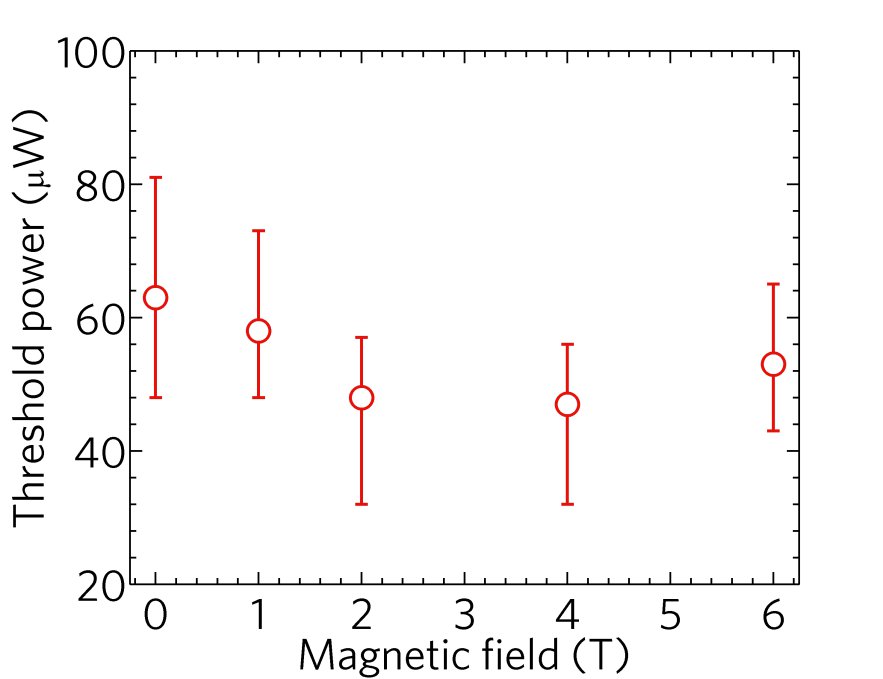}
\caption{Reduction of the threshold power in magnetic field.}
\label{im:Fig3}
\end{figure*}

The reduction of the threshold power in magnetic field is illustrated in Fig.~\ref{im:Fig3}. The origin of this effect is the reduction of the density of states at the ground state due to the energy splitting induced by magnetic field. This effect is  discussed in more detail in  \citep{Rousset_PRB2017}.

\begin{figure*}[h]
\centering
\includegraphics[width=.65\paperwidth]{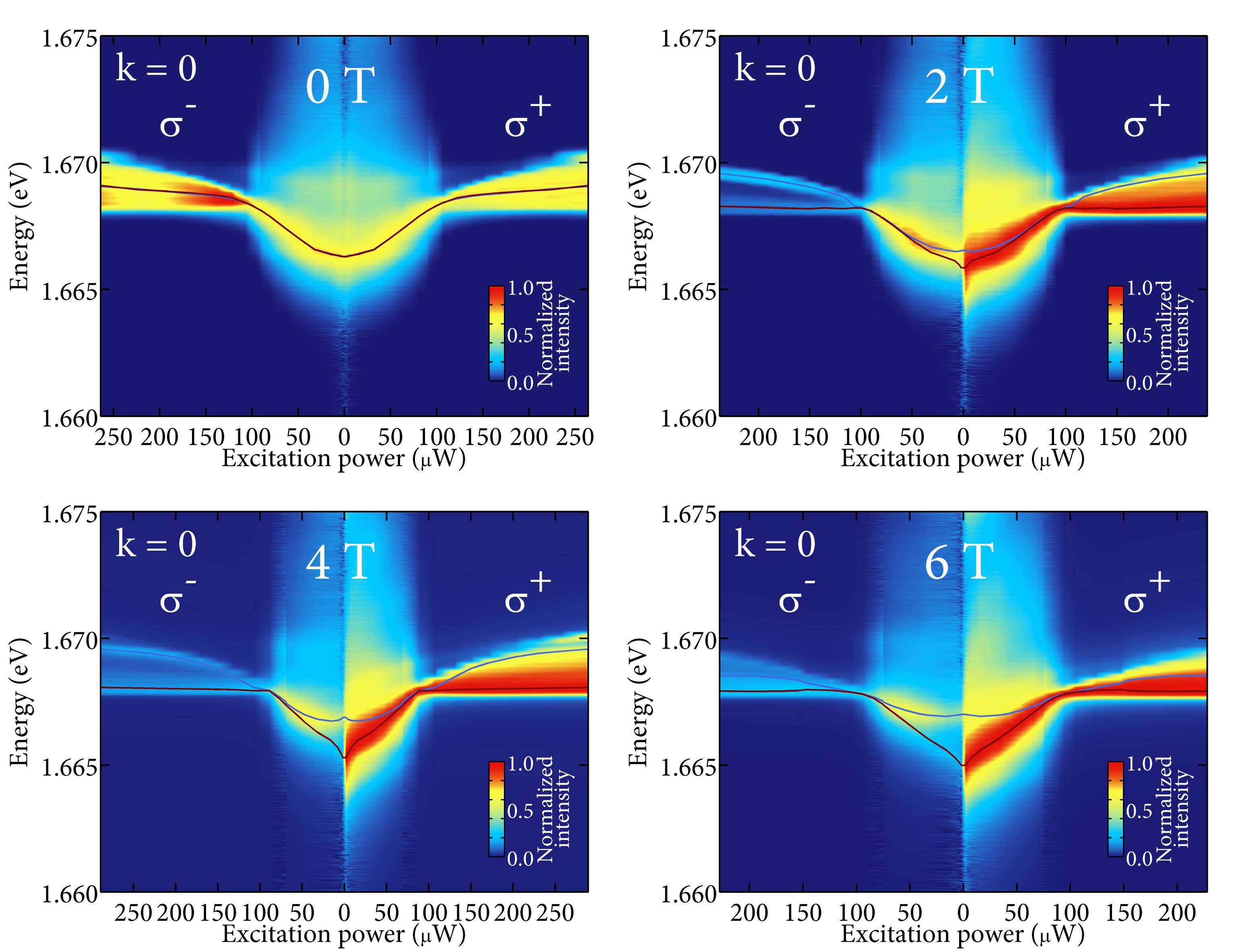}
\caption{Cross-section of the polariton emission at the normal incidence ($k=0$) for magnetic fields 0~T, 2~T, 4~T and 6~T. The solid lines mark the emission energy in $\sigma^+$ (red) and $\sigma^-$ (blue) polarizations.}
\label{im:Fig4}
\end{figure*}

To trace the condensate we plot the cross-section of Fig.~\ref{im:Fig1} at zero emission angle, where the condensate is visible. This is illustrated in Fig.~\ref{im:Fig4} for all measured magnetic fields. The polariton Zeeman splitting is visible at low excitation power as the energy difference between $\sigma^+$  and $\sigma^-$  polarized states as in Ref.~\citep{Mirek_PRB2017}.

\begin{figure*}[h]
\centering
\includegraphics[width=.6\paperwidth]{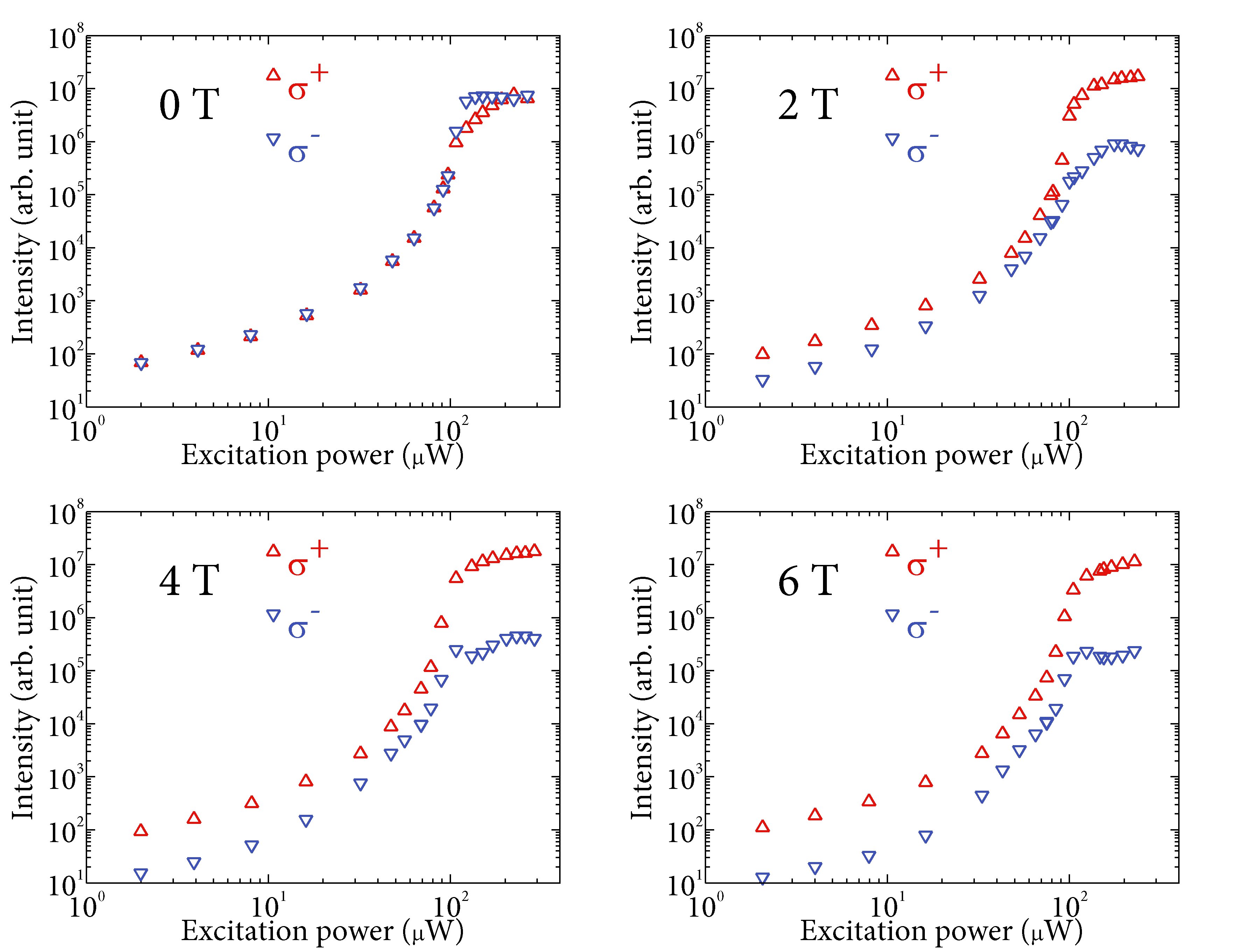}
\caption{The intensity of the signal at the bottom of lower polariton branch versus excitation power in magnetic field 0~T, 2~T, 4~T and 6~T. The condensation threshold at 0~T at 60~$\upmu$W and is slightly decreasing in magnetic field.}
\label{im:Fig5}
\end{figure*}

The accompanying change of the condensate emission intensity compared between $\sigma^+$ and $\sigma^-$ polarizations for each magnetic field is illustrated in Fig.~\ref{im:Fig5}. 

\section*{Excitonic reservoir}
\begin{figure*}[h]
\centering
\includegraphics[width=.6\paperwidth]{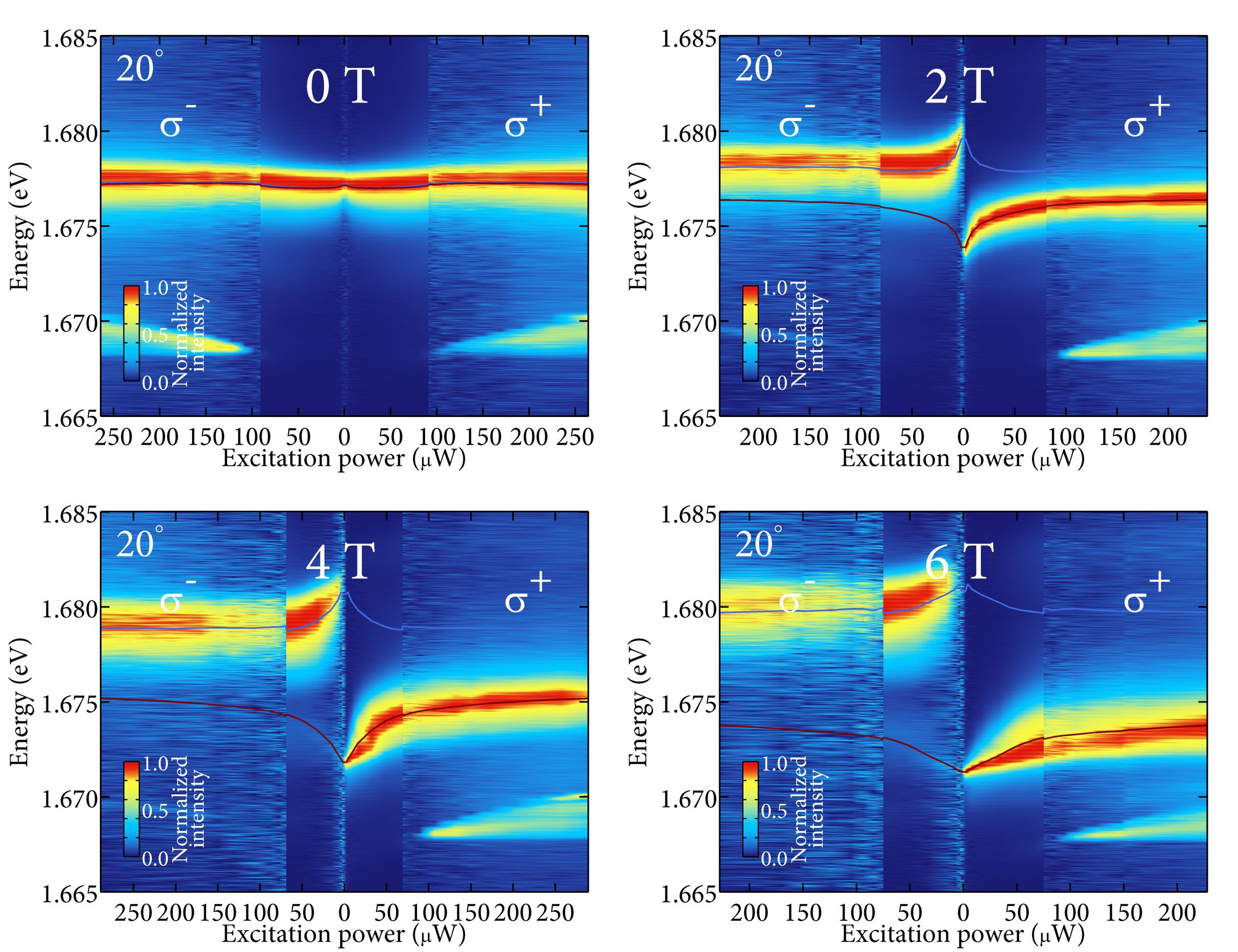}
\caption{Exciton-polariton emission at high angles illustrating the reduction of exciton Zeeman splitting due to the depolarization of Mn ions due to the high numbers of electrons and holes created in the system by high excitation power. The solid lines mark the emission energy in $\sigma^+$ (red) and $\sigma^-$ (blue) polarizations.}
\label{im:Fig6}
\end{figure*}

The cross-section of the polariton emission at high angles (20\,deg.) is illustrated in Fig.~\ref{im:Fig6}. With  increasing excitation power Zeeman splitting is suppressed, but even at 2\,T at highest excitation power is non-zero. Signal around 1.67\,eV is a residual emission  originating  from the condensate.

\section*{Theoretical modeling}

We describe the model that was used to obtain the theoretical results marked by solid lines in Fig.~2. in the main text. The phenomenological model of polariton dynamics includes polariton kinetics that are accounted for in a similar way as in previous works~\cite{Kavokin_PRL2004,Iorsh_PRB2012,Bhattacharya_PRB2016} in which spinful condensates were considered. Additionally, we describe the spin Meissner effect adapting the description formulated in~\citep{Rubo_2006,Shelykh_SuperlatticesMicrostruct2007} but generalised to the system in a nonequilibrium state. To account for both the reservoir dynamics and condensate formation, as well as the photon lasing at high pumping powers, we propose the equations for the reservoir density $n_{\rm R}(t)$ and its average spin ${\bf S}_{\rm R}(t)$, condensate density $n_{\rm c}(t)$, assuming that the spin of the condensate always corresponds to minimum of energy ${\bf S}_{\rm c}(t)={\bf S}^{\rm min}_{\rm c}(t)$, and lasing mode occupation $n_{\rm L}(t)$
\begin{align*}
\frac{\partial n_{\rm R}}{\partial t} &= P-\gamma_{\rm R} n_{\rm R}-R_{\rm sc} n_{\rm R} n_{\rm c}-R_{\rm L} n_{\rm R} n_{\rm L}, \\
\frac{\partial {\bf S}_{\rm R}}{\partial t} &= -\gamma_{\rm rel} \left({\bf S}_{\rm R}-{\bf S}_{\rm R}^{\rm min}\right)-R_{\rm sc} n_{\rm c} {\bf S}_{\rm R}-\gamma_{\rm R} {\bf S}_{\rm R}, \\
\frac{\partial n_{\rm c}}{\partial t} &= R_{\rm sc} n_{\rm R} n_{\rm c}-\gamma_{\rm c} n_{\rm c}, \\
\frac{\partial n_{\rm L}}{\partial t} &= R_{\rm L} n_{\rm R} n_{\rm L}-\gamma_{\rm L} n_{\rm L},
\end{align*}
where $P$ is the pumping rate, $\gamma_{\rm R, c}$ are excitonic and photonic decay rates, $R_{\rm sc} \approx R_{0} + R_1E_{\sigma_+}$ is the reservoir-condensate scattering rate, approximately linearly dependent on the condensate energy, $R_{\rm L}$ is rate of scattering into the lasing mode, $\gamma_{\rm rel} = \gamma_{\rm phonon} + \Gamma_{\rm int} n_{\rm R}$ is the reservoir relaxation rate, which contains phonon mediated relaxation $\gamma_{\rm phonon}$ and interaction mediated relaxation $\Gamma_{\rm int}$. 
The reservoir and condensate spins that minimize the energy of the system  ${\bf S}_{\rm R,c}^{\rm min}$ are calculated as in \citep{Rubo_2006}, and \citep{Shelykh_SuperlatticesMicrostruct2007}, from the minimization of the free energy in the simplified form
\begin{equation}\label{En}
F=-2\Omega_z S^z+2\Omega_x S^x+\frac{\alpha}{2} \left(n_+^2 + n_-^2\right) - \mu n,
\end{equation}
where $\Omega_z$ is the excitonic giant Zeeman splitting due to the magnetic field parallel to the sample growth axis,  $\Omega_x$ is the linear polarization splitting due to the sample anisotropy or the residual transverse magnetic field, $S^{x,z}=S_R^{x,z} + X S_c^{x,z}$ is the $z$ component of the total excitonic spin, where $X$ is the Hopfield coefficient of condensed polaritons, $n^\pm=n_{\rm R}^\pm + X n_{\rm c}^\pm$ where $n_{\rm R, c}^{\pm}=(n_{\rm R, c}/2)  \pm S^z_{\rm R,c}$, $\alpha$ is the parallel spin interaction coefficient, and we neglect the interaction between polaritons with antiparallel $z$-spin projection.

The energy and polarization of emission lines are calculated by variation $\delta F$ of the free energy~(\ref{En}) with respect to single-particle excitations ($\delta n = 2 |\delta {\bf S}_{R,c}| = 1 \ll n$). The polariton eigenmodes correspond to extrema of the free energy.
For instance, if $\Omega_x S^x\ll \Omega_z S^z$ (which is fulfilled in most cases), the energy of $\sigma_\pm$ excitations is
\begin{equation} \label{En_exc}
E_{\sigma_\pm} = E_0+X \left(\mp \Omega_z+\alpha \left(n_{\rm R}^{\pm}+X n_{\rm c}^{\pm}\right)\right),
\end{equation}
where $E_0$ is a constant energy shift treated as a fitting parameter.
The above approach gives results identical as the Bogoliubov method from \cite{Rubo_2006} taken at $k_{\perp}=0$, but is valid also in the non-condensed case. In the case of a condensed system, emission from both the condensate and the reservoir can be calculated by appropriate variations of condensate or reservoir density and spin.


We find stationary states of the above set of equations and compare them to the experimental results in Fig.~2 in the main text. Although system is not stationary in the case of pulsed pumping, the agreement is very good on a quantitative level. We explain this by the existence of a relatively long-lived inactive reservoir of high-energy excitons or free carriers, which leads to a long timescale of the effective condensate pumping, which drives the system close to a steady state even in the case of pulsed pumping.

The following values of dimensionless model parameters were found by fitting to experimental data: $\gamma_{\rm phonon}=0.4,\Gamma_{\rm int}=2,\gamma_{\rm R}=1,\gamma_{\rm c}=1.7,\alpha=1, R_0=0.05, R_1=1.8, R_{\rm L}=1, E_{\rm L}=0.8879$, $\Omega_x=0.2$, $\Omega_z=1$ at magnetic field strength $6\,$T. The numerical results were scaled to physical units by linear rescaling of the energy and pumping power. We note that these theoretical fits are not sufficient to determine values of physical parameters of the model. For instance, ratios such as $\gamma_{\rm c}/\gamma_{\rm R}$ or $R_1/R_0$  have physical significance, but since we are looking for steady states only, it is not possible to determine the absolute values of these parameters. On the other hand, due to the scaling properties of the model and the unknown value of density or spin density it is not possible to determine ratios of parameters $\gamma_i$ to $R_j$ or $\Omega_i$ to $\alpha$. Consequently, any attempt to determine absolute values of physical parameters would be highly speculative.

To take into account the effect of photon lasing at very high pumping intensity, we assume that in the regime when number of photons $n_L$ is much larger than the number of polaritons in the condensate $n_C$,  the energy of the $\sigma^+$ polariton line is equal to $E_{\sigma_+} = E_{\rm L}$, where $E_L$ is the cavity photon energy, and emission is fully $\sigma^+$ polarized. 

\bibliography{bib_GiantSpinnMeissner}
\bibliographystyle{apsrev4-1}